# A clustering-based data reduction for very large spatio-temporal datasets

Nhien-An Le-Khac[1], Martin Bue[2], Michael Whelan[1], M-Tahar Kechadi[1],

[1] School of Computer Science and Informatics, University College Dublin, Belfield, Dublin 4, Ireland
[2] Ecole Polytechnique Universitaire de Lille, Villeneuve d'Ascq cedex, France

[1]{an.lekhac, michael.whelan,tahar.kechadi}@ucd.ie
[2]Martin.Bue@polytech-lille.net

**Abstract.** Today, huge amounts of data are being collected with spatial and temporal components from sources such as meteorological, satellite imagery etc. Efficient visualisation as well as discovery of useful knowledge from these datasets is therefore very challenging and becoming a massive economic need. Data Mining has emerged as the technology to discover hidden knowledge in very large amounts of data. Furthermore, data mining techniques could be applied to decrease the large size of raw data by retrieving its useful knowledge as representatives. As a consequence, instead of dealing with a large size of raw data, we can use these representatives to visualise or to analyse without losing important information. This paper presents a new approach based on different clustering techniques for data reduction to help analyse very large spatio-temporal data. We also present and discuss preliminary results of this approach.

## 1   Introduction

Many natural phenomena present intrinsic spatial and temporal characteristics. Besides traditional applications, recent concerns about climate change, the threat of pandemic diseases, and the monitoring of terrorist movements are some of the newest reasons why the analysis of spatio-temporal data has attracted increasing interest. With the recent advances in hardware, high-resolution spatio-temporal datasets are collected and stored to study important changes over time, and patterns of specific events. However, these datasets are often very large and grow at a rapid rate. So, it becomes important to be able to analyse, discover new patterns and trends, and display the results in an efficient and effective way.

Spatio-temporal datasets are often very large and difficult to analyse [1][2][3]. Fundamentally, visualisation techniques are widely recognised to be powerful in analysing these datasets [4], since they take advantage of human abilities to perceive visual patterns and to interpret them [5]. However spatial visualisation techniques currently provided in the existing geographical applications are not adequate for decision-support systems when used alone. For instance, the problems of how to visualise the spatio-temporal multi-dimensional datasets and how to define effective visual interfaces for viewing and manipulating the geometrical components of the spatial data [6] are the challenges. Hence, alternative solutions have to be defined. Indeed, new solutions should not only include a static graphical view of the results produced during the data mining (DM) process, but also the possibility to dynamically and interactively obtain different spatial and temporal views as well as interact in different ways with them. DM techniques have been proven to be of significant value for analysing spatio-temporal datasets [7][8]. It is a user-centric, interactive process, where DM experts and domain experts work closely together to gain insight on a given problem. In particular, spatio-temporal data mining is an emerging research area, encompassing a set of exploratory, computational and interactive approaches for analysing very large spatial and spatio-temporal datasets. However, several open issues have been identified ranging from the definition of techniques capable of dealing with the huge amounts of spatio-temporal datasets to the development of effective methods for interpreting and presenting the final results.

Analysing a database of even a few gigabytes is an arduous task for machine learning techniques and requires advanced parallel hardware and algorithms. Huge datasets create combinatorially explosive search spaces for DM algorithms which may make the process of extracting useful knowledge infeasible owing to space and time constraints. An approach for dealing with the intractable problem of learning from huge databases is to select a small subset of data for mining [2]. It would be convenient if large databases could be replaced by a small subset of representative patterns so that the accuracy of estimates (e.g., of probability density, dependencies, class boundaries) obtained from such a reduced set should be comparable to that obtained using the entire dataset.

Traditionally, the concept of data reduction has received several names, e.g. editing, condensing, filtering, thinning, etc, depending on the objective of the reduction task. Data reduction techniques can be applied to obtain a reduced representation of the dataset that is much smaller in volume, yet closely maintains the integrity of the original data. That is, mining on the reduced dataset should be more efficient yet produce the same analytical results. There has been a lot of research into different techniques for the data reduction task which has lead to two different approaches depending on the overall objectives. The first one is to reduce the quantity of instances, while the second is to select a subset of features from the available ones.

In this paper, we will focus on the first approach to data reduction which deals with the reduction of the number of instances in the dataset. This approach can be viewed as similar to sampling, a technique that is commonly used for selecting a subset of data objects to be analysed. There are different techniques for this approach such as the scaling by factor [9][10], data compression [11], clustering [12], etc.

Often called numerosity reduction or prototype selection, instance reduction algorithms are based on a distance calculation between instances in the dataset. In such cases selected instances, which are situated close to the centre of clusters of similar instances, serve as the reference instances. In this paper we focus on spatio-temporal clustering technique. Clustering is one of the fundamental techniques in DM [2]. It groups data objects based on information found in the data that describes the objects and their relationships. The goal is to optimise similarity within a group of objects and the dissimilarity between the groups in order to identify interesting structures in the underlying data. Clustering is used on spatio-temporal data to take advantage of the fact that, objects that are close together in space and/or in time can usually be grouped together. As a consequence, instead of dealing with a large size of raw data, we can use these cluster representatives to visualise or to analyse without losing important information.

The rest of the paper is organised as follows. In Section II we discuss background and related work. Section III describes in detail our data reduction technique based on clustering. Section IV we evaluate the results of our data reduction technique as a pre-processing step on a very large spatio-temporal dataset. In Section V we discuss future work and conclude.

## 2 Background

### 2.1 Data Reduction

**Sampling.** The simplest approach for data reduction, the idea is to draw the desired number of random samples from the entire dataset. Various random, deterministic and density biased sampling strategies exist in literature [7][13]. However, naive sampling methods are not suitable for real world problems with noisy data, since the performance of the algorithms may change unpredictably and significantly. The random sampling approach effectively ignores all the information present in the samples not chosen for membership in the reduced subset. An advanced data reduction algorithm should include information from all samples in the reduction process [14][15].

**Discretisation**. Data discretisation techniques can be used to reduce the number of values for a given continuous attribute by dividing the range of the attribute into intervals. Interval labels can then be used to replace actual data values. Replacing numerous values of a continuous attribute by a small number of interval labels thereby reduces and simplifies the original data. In [9] the data reduction consists of discretising numeric data into ordinal categories. The process starts by first being given a number of points (called split points or cut points) to split the entire attribute range, and then repeats this recursively on the resulting intervals. The problem with this approach taken in [9] is that the interval selection process drops a large percentage of the data while trying to reduce the range of values a dimension can have.

### 2.2 Spatio-Temporal Data Mining

Spatio-temporal DM represents the junction of several research areas including machine learning, information theory, statistics, databases, and geographic visualisation. It includes a set of exploratory, computational and interactive approaches for analysing very large spatial and spatio-temporal datasets. Recently various projects have been initiated in this area ranging from formal models [4][16] to the study of the spatio-temporal data mining applications [5][16]. In spatio-temporal data mining the two dimensions "spatial" and "temporal" have added substantial complexity to the traditional DM process. It is worth noting, while the modelling of spatio-temporal data at different levels of details presents many advantages for both the application and the system. However it is still a challenging problem. Some research has been conducted to integrate the automatic zooming of spatial information and the development of multi-representation spatio-temporal systems [17][18][19]. However, the huge size of datasets is an issue with these approaches. In [8][9], the authors proposed a strategy that is to be incorporated in a system of exploratory spatio-temporal data mining, to improve its performance on very large spatio-temporal datasets. This system provides a DM engine that can integrate different DM algorithms and two complementary 3-D visualisation tools. This approach reduces their datasets by scaling them by a factor F; it simply runs through the whole dataset taking one average value for the $F^3$ points inside each cube of edge F. This reducing technique has been found to be inefficient as a data reduction method which may lose a lot of important information contained in the raw data.

## 2.3 Related Work

To the best of our knowledge, there is only our recent work [12] which proposed a knowledge-based data reduction method. This method is based on clustering [1] to cope with the huge size of spatio-temporal datasets in order to facilitate the mining of these datasets. The main idea is to reduce the size of that data by producing a smaller representation of the dataset, as opposed to compressing the data and then uncompressing it later for reuse. The reason is that we want to reduce and transform the data so that it can be managed and mined interactively. Clustering technique used in this approach is K-Medoids [12]. The advantage of this technique is simple; its representatives (medoids points) cannot however reflect adequately all important features of the datasets. The reason is that this technique is not sensitive to the shape of the datasets (convex).

## 3 Knowledge-based data reduction

In this section, we present a new approach to improve our previous data reducing method [12]. We summarise firstly a framework of spatio-temporal data mining where our reducing method will be applied. Next, we describe our knowledge-based approach in an analytical way with an algorithm and the discussion on its issues.

### 3.1 Spatio-temporal data mining framework

As described in [8][9], our spatio-temporal data mining framework consists of two layers: mining and visualisation. The mining layer implements a mining process along with the data preparation and interpretation steps. For instance, the data may need some cleaning and transformation according to possible constraints imposed by specific tools, algorithms, or users. The interpretation step consists of using the selected models returned during the mining to effectively study the application's behaviour. The visualisation layer contains different visualisation tools that provide complementary functionality to visualise and interpret mined results. More details on the visualisation tools can be found in [8][9].

In the first layer, we applied the two-pass strategy. The reason is that the raw spatial-temporal dataset is too large for any algorithm to process; the goal of this strategy is to reduce the size of that data by producing a smaller representation of the dataset, as opposed to compressing the data and then uncompressing it later for reuse. Furthermore, we aim to reduce and transform the data so that it can be managed and mined interactively. In the first pass, the data objects are grouped according to their close similarity and then these groups are analysed by using different DM techniques based on specific objectives in the second pass. The objective of the first pass is to reduce the size of the initial data without losing any relevant information. On the other hand, the purpose of the second pass is to apply mining technique such as clustering, association rules on the tightly grouped data objects to produce new knowledge and ready for evaluation and interpretation. In the first implementation of this strategy, a scaling-based approach was used for the first past. Although it is simple and easy to implement, it loses a lot of important information contained in the raw data.

### 3.2 Clustering for data reduction

In this paper, we propose a new data reduction method based on clustering to help with the mining of the very large spatio-temporal dataset. Clustering is one of the fundamental techniques in DM. It groups data objects based on the characteristics of the objects and their relationships. It aims at maximising the similarity within a group of objects and the dissimilarity between the groups in order to identify interesting structures in the underlying data. Some of the benefits of using clustering techniques to analyse spatio-temporal datasets include a) the visualisation of clusters can help with understanding the structure of spatio-temporal datasets, b) the use of simplistic similarity measures to overcome the complexity of the datasets including the number of attributes, and c) the use of cluster representatives to help filter (reduce) datasets without losing important/interesting information.

We have implemented a combination of density-based and graph-based clustering in our approach. We have chosen a density-based method rather than other clustering method such as centre-based because it is efficient with spatial datasets as it takes into account the shape (convex) of the data objects [1]. However, it would be a performance issues when a simple density-based algorithm applied on huge amount of spatial datasets including differences in density. Indeed, the running times as well as the choice of suitable parameters are performance impacts of complex density-based algorithms [2]. In our algorithm, a modification of DBSCAN is used because DBSCAN algorithm [20] is simple; it is also one of the most efficient density-based algorithms, applied not only in research but also in real applications.

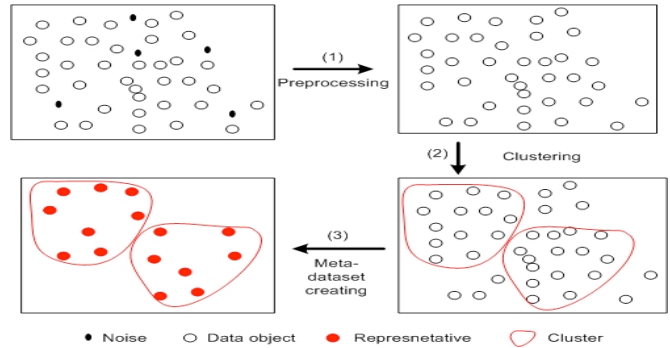

**Fig. 1.** Step by step view of the first pass of the strategy.

In order to cope with the problem of differences in density, we combine DBSCAN with a graph-based clustering algorithm. Concretely, the Shared Nearest Neighbor Similarity (SNN) algorithm [21] is used to firstly build a similarity graph. Next, DBSCAN will be carried out based on the similarity degree. The advantage of SNN is that it address the problems of low similarity and differences in density. Another approach with a combination of SNN and DBSCAN was proposed in [2]. However, it is not in the context of data reduction and it did not take into account the problem of the huge size of the datasets in the context of memory constraint.

Fig.1 shows an overview of our reduction method including four steps: (1) a simple pre-processing is applied on raw datasets to filter NULL values. (2) SNN similarity graph is built for all datasets. Similarity degree of each data object is also computed in this step. The two parameters *Eps* and *Minpts* are selected based on these similarity degrees. Then, DBSCAN-based algorithm is carried out on the datasets to determine *core objects*, *specific core objects*, *density-reachable objects*, *density-connected objects*. These objects are defined in [20][22].

Clusters are also built based on *core objects* and *density-reachable* features in this step. Data objects which do not belong to any cluster will be considered as noise objects. *Core objects* or *specific core objects* are selected as cluster representatives that form a new (meta-) dataset (3). This dataset can then be analysed and produce useful information (i.e. models, patterns, rules, etc.) by applying other DM techniques (second pass of our framework). It is important to note that data objects that have a very high similarity between each other can be grouped together in the same clusters. As a result of this pass, the new dataset is much smaller than the original data without losing any important information from the data that could have an adverse effect on the result obtained from mining the data at a later stage.

## 4  Evaluation and Analysis

In this section, we study the feasibility of data reducing for spatio-temporal datasets by using DM techniques described in Section III. We compare also this approach with two other ones: scaling [8] (cf. 2.2) and K-Medoids clustering [2]. The dataset is the

Isabel hurricane data [23] produced by the US National Centre for Atmospheric Research (NCAR). It covers a period of 48 hours (time-steps). Each time-step contains several atmospheric variables. The grid resolution is 500×500×100. The total size of all files is more than 60GB (~1.25 GB for each time-step). The experimentation details and a discussion are given below.

### 4.1 Experiments

The platform of our experimentation is a PC of 3.4 GHz Dual Core CPU, 3GB RAM using Java 1.6 on Linux kernel 2.6. Datasets of each time-step include 13 non-spatio attributes, so-called dimensions. In this evaluation, QCLOUD is chosen for analysis; it is the weight of the cloud water measured at each point of the grid. The range of QCLOUD value is [0…0.00332]. Different time-steps are chosen to evaluate: 2, 18, and 42. Totally, the testing dataset contains around 25 million data points of 4 dimensions X, Y, Z, QCLOUD for each time step.

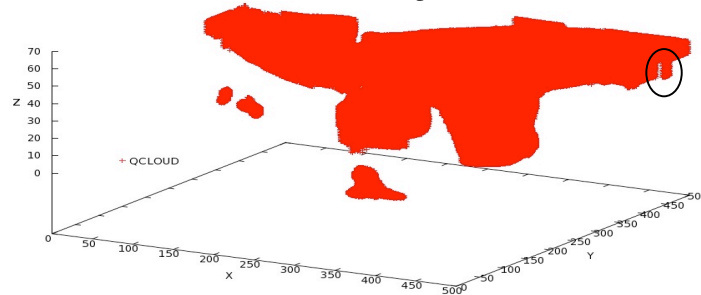

(a) Time-step 2

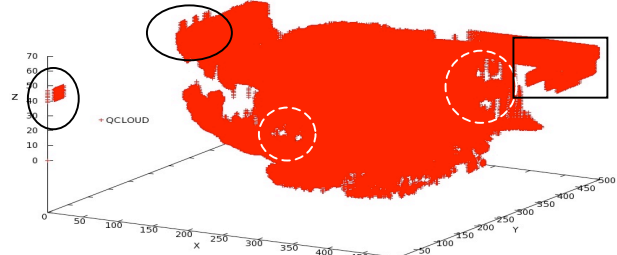

(b) Time-step 18

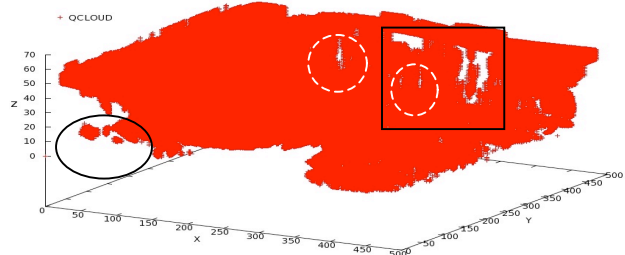

(c) Time-step 42

**Fig.2.** All dataset.

Fig.2 shows all data points before processing by any data reduction technique for testing dataset. Fig.3 shows the scaling results of our testing dataset in the grid coordinate at the selected time-steps. The scaling factor chosen is 50 ×50 × 50 for X, Y, Z i.e. we obtain 125000 data points after scaling as representative points. Fig.4 shows the results of our density-based clustering approach presented in Section 3 on the testing dataset. We also show representatives (*specific core point*) of each cluster. The number of representatives is approximate 120000. Fig.5 shows our testing dataset after the reducing process by a K-Medoids clustering. The number of clusters is 2000. We only show 50 data points including the medoid point of each cluster as representatives. We have totally 100000 data points for this case.

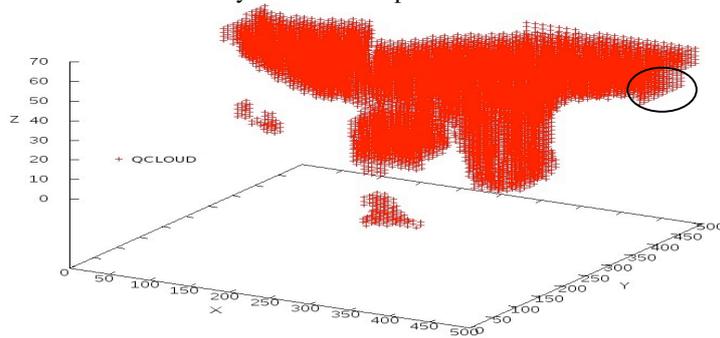

(a) Time-step 2

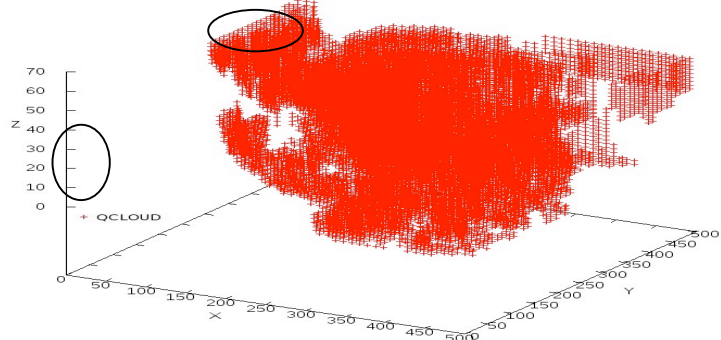

(b) Time-step 18

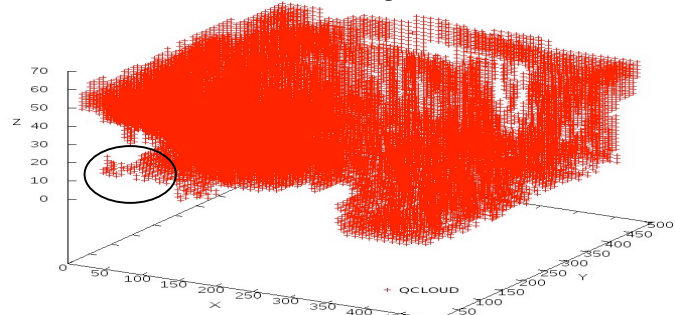

(c) Time-step 42

**Fig.3.** Scaling by 50 × 50 × 50.

## 4.2 Analysis

As shown in Fig. 3, 4 and 5, representative points (or representative, in brief) could reflect the general shape of hurricane based on (X,Y,Z,QCLOUD) comparing to their whole original points (Fig.2) in different time-steps. By observing these figures, we recognise that the scaling approach cannot reflect the border of data points in details. For instance, it cannot show the shape of data points in the upper left corner (Fig.2a, Fig.3a and Fig.4a, data points in the circle). The reason is that in the scaling approach, all data points in a cube are reduced to the centre point. If this centre point is outside the dense areas of this cube then it cannot exactly represent the dense feature of this cube. Moreover, sometimes this position may get Null value while other positions in this cube are not null.

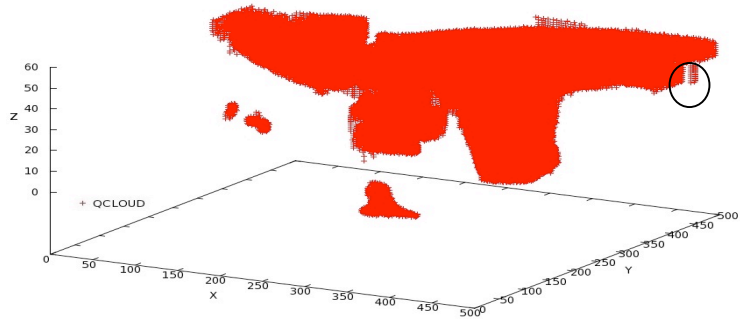

(a) Time-step 2

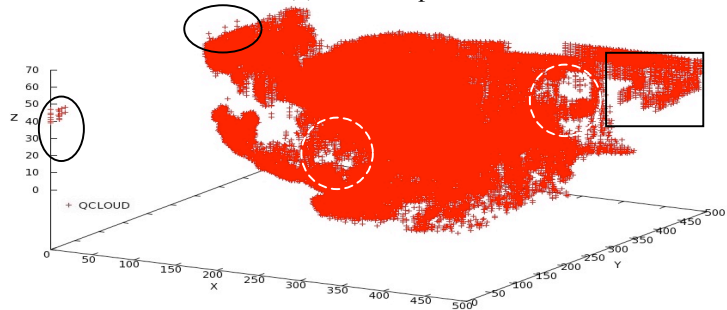

(b) Time-step 18

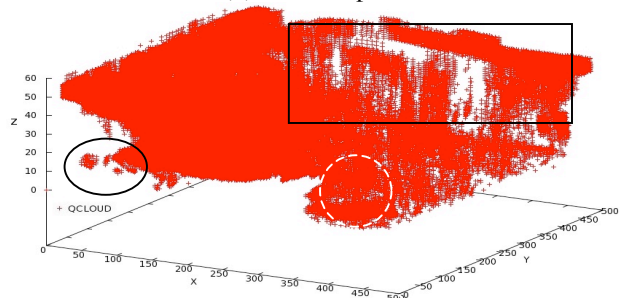

(c) Time-step 42

**Fig.4.** SNN-DBSCAN-based.

Fig.2b, Fig.3b and Fig.4b for the time-step 18 also confirm our observation. The scaling approach (Fig.3b) cannot show the data points on the extreme left of figure (near the Z-axis). Other circles in Fig2, Fig.3 and Fig.4 remark positions where this approach cannot represent efficiently. Indeed, the density-based approach (Fig.4) can show different holes (remark by dash circle in Fig.4 and Fig.2) clearer than the original ones (Fig.2) because in our approach, redundant data points are eliminated. Note that these holes are normally very important to geography experts to study the important features of a hurricane. They are sometimes not clear in early hours of the hurricane (Fig.2a, Fig.4a) Furthermore, representatives in our approach form different clusters and further analysis can base on this feature to extract more hidden knowledge. It is clearly that we can significantly reduce running time in following stages of spatio-temporal data mining because the size of representatives is less than 10% of all dataset. Besides, these representatives also reflect the movement of hurricane as their whole origin points do.

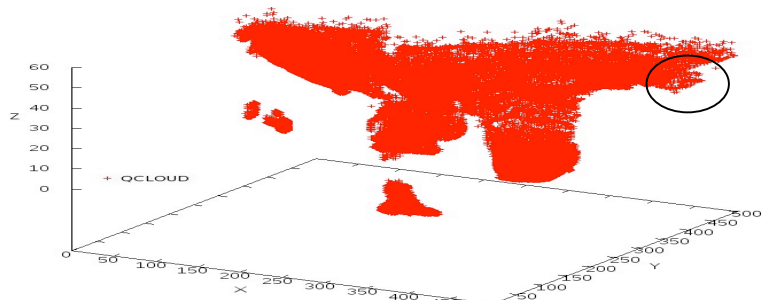

(a) Time-step 2

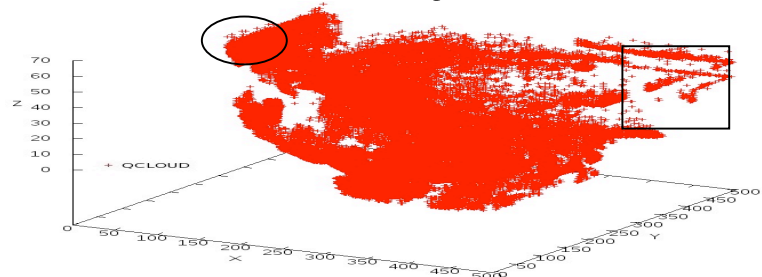

(b) Time-step 18

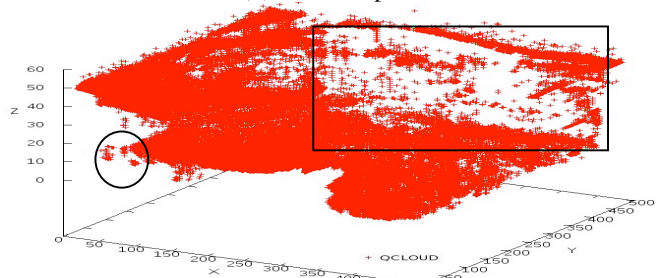

(c) Time-step 42

**Fig.5.** K-Medoids.

In our approach, specific core points are used instead of core points. The reason is that the number of core points is approximately 90% of all dataset in our experiments. Thus, the reduction does not gain in term of size of dataset. On the other hand, if there is a limitation of main memory, then the multi-partition approach will be applied. For instance, we carried out our approach with 10 parts in a limited memory and it gives an approximate result in representatives comparing to the case where whole dataset loaded in memory. The reason is that we combine SNN algorithm with our approach and it can deal with the different densities in the datasets. Consequently, it can reduce the effect of the partition of the dataset. However, running time as well as choosing efficient parameters (e.g. number of partitions) for this approach is also a performance issue. Besides, K-medoid approach gives a better performance comparing to the scaling one (Fig.3,5 e.g. data points in the circle). However, it cannot efficiently reflect the shape of data comparing to the SNN-DBSCAN based (Fig.4,5 e.g. data points in the rectangular). As a brief conclusion, these experiments above show that with our new approach, the use of simple DM techniques can be applied to reduce the large size of spatio-temporal datasets and preserve their important knowledge used by experts.

## 5   Conclusion and Future work

In this paper, we study the feasibility of using DM techniques for reducing the large size of spatio-temporal datasets. As there are many reducing techniques presented in literature such as sampling, data compression, scaling, etc., most of them are concerned with reducing the dataset size without paying attention to their geographic properties. Hence, we propose to apply a clustering technique to reduce the large size without losing important information. We apply a density-based clustering that is a combination of SNN and DBSCAN-based algorithm on different time-steps. The experimental results show that knowledge extracted from mining process can be used as efficient representatives of huge datasets. Furthermore, we do not lose any important information from the data that could have an adverse effect on the result obtained from mining the data at a later stage. Besides, a solution for the limitation of memory is also proposed. We have reported some of these preliminary visual results for QCLOUD in its space X,Y,Z for three different time-steps (2, 18 and 42).

A more extensive evaluation is on-going. In the future we intend to analyse different combinations of dimensions over more time steps to try and find hidden information on their relationships with each other. Indeed, we are currently testing with a hybrid approach where density-based and centre-based clusterings are used to increase the performance in terms of running time and representative positions.